\documentstyle[aps,twocolumn,prb,epsf]{revtex} 

\begin{document}

\draft
\twocolumn[    
\hsize\textwidth\columnwidth\hsize\csname @twocolumnfalse\endcsname    

\title{Nonconstant electronic density of states
tunneling inversion for A15 superconductors: Nb$_3$Sn
}
\author{J. K. Freericks, Amy Y. Liu, A. Quandt}
\address{Department of Physics, Georgetown University, 
  Washington, DC 20057-0995, U.S.A.}
\author{J. Geerk}
\address{Forschungszentrum Karlsruhe, Institut f\"ur Festk\"orperphysik,
P.O. Box 3640, D-76021 Karlsruhe, Germany}
\date{\today}
\maketitle

\widetext
\begin{abstract}
We re-examine the tunneling data on A15 superconductors by performing 
a generalized McMillan-Rowell tunneling inversion that incorporates
a nonconstant electronic density of states obtained from band-structure
calculations.  For Nb$_3$Sn, we find that the fit to the experimental data 
can be slightly improved by taking into account the sharp structure in the 
density of states, but it is likely that such an analysis
alone is not enough to completely explain the superconducting tunneling
characteristics of this material.  Nevertheless, the extracted Eliashberg
function displays a number of features expected to be present for the 
highest quality Nb$_3$Sn samples.
\end{abstract}

]      

\narrowtext
\section{Introduction}

Twenty years ago, the A15 superconductors in the $A_3$B structure, with A a
transition metal and B an sp-metal, were the highest transition 
temperature superconductors known. The first such superconductor found
was V$_3$Si,
which was discovered\cite{v3si} in the 1950's to have a $T_c$ 
of about 17~K. In the ensuing years, other equilibrium compounds like
V$_3$Ga and Nb$_3$Sn were discovered to have  
$T_c$'s in the range of 15-18~K. 
B-element-poor compounds like
Nb$_3$Al, Nb$_3$Ga, Nb$_3$Ge, and Nb$_3$Si were also found, with $T_c$'s
as high as 23~K.  
The structural, electronic, magnetic, elastic, vibrational, and superconducting 
properties of these materials were widely studied.\cite{reviews,tunneling}

The use of tunneling spectroscopy to probe superconducting properties 
of the A15 materials was hindered by a number of materials issues.
The fabrication of high-quality tunnel junctions was difficult 
because the use of native oxides
for the tunneling barrier did not yield reproducible results, and hence
artificial barrier layers had to be grown on
top of the A15 superconductors.\cite{rudman,geerk}
As the quality of the tunneling data improved, it became 
clear that these materials do not exhibit the simple behavior seen in 
conventional
strong-coupling s-wave superconductors like Pb, Hg, and Nb.  In particular,
the reduced tunneling density of states
displays a rapid overswing\cite{mitrovic-over} followed by a sharp return 
to zero at energies near and above the maximum phonon energy.
A revision of the McMillan-Rowell 
tunneling analysis\cite{mcmillan-rowell} by Arnold and Wolf\cite{arnold}
allowed this data to be fit to high
precision by assuming the presence  of an additional normal-metal layer 
(characterized by a width and a scattering strength) 
between the superconductor and the insulating
barrier.   However, to fit the most 
recent experimental data\cite{geerk} within this proximity-effect modified 
tunneling theory, one needs to assume that the width of the normal region
approaches zero while its scattering strength becomes unusually
large. 

Alternatively, it has been suggested that the rapid overswing in
the high-energy regime in the
tunneling density of states is related to the presence of
sharp structures in the electronic
density of states near the Fermi level.\cite{mitrovic-over}  
Band-structure calculations\cite{band-general}
show that the electronic
density of states in these materials has peaks of width on the order of
100~meV near the Fermi energy.  Such sharp structures require, at
the very least, a reformulation of the Migdal-Eliashberg\cite{migdal-eliashberg}
many-body analysis to include effects
of a nonconstant electronic density of states\cite{kr,pickett,mitrovic}
within an energy range on the
order of the maximal phonon energies of the material.

Other theories of superconductivity in A15 
compounds go beyond simply generalizing the standard
Migdal-Eliashberg theory to allow for 
energy dependence in the density of states. 
For example, Yu and Anderson\cite{yu}
examine what happens in an electron-phonon
system that is coupled strongly enough to have the single-electron (polaronic)
phase become unstable to bipolaronic (pre-formed pair) phases.  Such systems
display quite different behavior, but these theories have not been 
developed to the point where direct comparison to materials-specific 
tunneling conductances can be made. 

Since the discovery of the high-$T_c$ cuprates,  work on the A15 compounds
has virtually ceased - this despite the many fundamental
questions about these materials that remain open.
In this paper,  we investigate whether high-quality tunneling data
for the A15 materials can be understood within a 
conventional Migdal-Eliashberg framework generalized to include
an energy-dependent electronic density of states. 
We perform
this analysis and extract an experimentally fit Eliashberg
function $\alpha^2F(\Omega)$, and Coulomb pseudopotential $\mu^*$,
for Nb$_3$Sn, which has $T_c=18$~K.  Data from high-quality tunnel 
junctions grown on this material are available.\cite{geerk}

In Section II, we derive the formalism needed to numerically perform the
tunneling inversion including a nonconstant electronic density of states.
This derivation and the computational algorithm are different from those
developed previously in that an exact analytic continuation
that properly treats the Coulomb pseudopotential and allows the 
calculations to be performed relative to the normal state is employed.  
In Section III, we present our
numerical results for the tunneling inversion with both constant and nonconstant
electronic density of states.  Conclusions are presented in Section IV.

\section{Formalism}

Our analysis begins with the calculation of the band structure 
and electronic density of states (DOS).  Calculations are carried out 
using the VASP package,\cite{vasp}
a plane-wave based density-functional code using ultrasoft
pseudopotentials.\cite{pseudo}
The electronic wave functions are expanded in  plane waves up to a 
cutoff of 219 eV, and the electron-electron
interaction is treated within the local density approximation (LDA) with  the
Ceperley-Alder exchange-correlation functional.\cite{exch-co}
The Brillouin zone is sampled on a Monkhorst-Pack mesh\cite{monkhorst-pack}
of at least $20\times 20\times 20$
points.  We find the optimized crystal structure to be tetragonal,
with a small distortion of the Nb sublattice.\cite{lu-klein} 
The peaks in the density of states near the Fermi level are  very
sensitive to this sublattice distortion.

In conventional Migdal-Eliashberg theory,\cite{migdal-eliashberg} the
electronic density of states is chosen to be a constant (with an infinite
``bandwidth''), and the energy cutoff of the theory is provided by the
finite range of the Eliashberg function $\alpha^2F(\Omega)$, which measures
the ability of a phonon of energy $\Omega$ to scatter electrons on the
Fermi surface.  When performing the many-body theory calculations, one
begins on the imaginary axis, where the Coulomb pseudopotential
has a sharp cutoff,\cite{leavens-fenton}  and then performs
an exact numerical
analytic continuation\cite{marsiglio} to calculate real-axis
properties.  This technique allows for a proper treatment of the ``soft''
cutoff for the Coulomb pseudopotential on the real axis.  In addition,
because the superconducting and normal states appear more and more
similar at high energies, and because of the finite frequency cutoff (chosen
to be 6$\Omega_{\rm max}$ here), the self-energy begins to deviate from
the exact result in the normal state as the energy increases.  One can
correct for this by performing the perturbative expansion relative to
the normal state.  In this case, one adds the exact normal-state
self-energy minus the normal-state self-energy calculated with the energy 
cutoff used in the superconducting theory.\cite{marsiglio-correction}
Such a scheme was used when examining effects of vertex corrections\cite{pb_us}
in Pb.  When an energy-dependent electronic
density of states is used,  such a computational scheme becomes 
problematic because there is no longer an exact analytic expression for the 
normal-state
self-energy.  Furthermore, a choice must be made for the energy
cutoff, because, assuming a bandwidth on the order of a few eV and a 
temperature on the order of 0.1~meV, the number of Matsubara frequencies
required for an energy cutoff set by the bandwidth
would be too large to perform calculations efficiently.

We adopt an alternative scheme here.  We start by first calculating the
normal-state self-energy at $T=0$.  Since the Matsubara frequencies become 
a continuum at $T=0$, such a calculation can be performed simply by
replacing the Matsubara summations by integrals along the imaginary
axis, which are computed using conventional quadrature techniques.  Next,
we choose the energy cutoff to be the same as that used in the constant
density of states calculations, namely 6$\Omega_{\rm max}$.  We calculate
the self-energy in the normal state  at $T$ (relative to the normal
state at $T=0$) with the
finite energy cutoff, and then add the $T=0$
normal-state self-energy (with no cutoff)
to the finite-temperature self-energy. Then we calculate the superconducting
self-energy relative to the normal state at $T$ using the same cutoff.
Finally, we add the full normal-state self-energy at $T$ to get the
superconducting self-energy.

Our many-body formalism on the imaginary axis follows most closely to
Ref.~\onlinecite{mitrovic}.  We use a Nambu-Gor'kov formalism and evaluate
the self-consistent perturbation theory (using dressed phonons) in the
Hartree-Fock approximation. (The Hartree term provides just a constant shift
to the chemical potential and is ignored.)  Perturbative calculations are
performed for the normal state and the superconducting state.  We begin
our analysis in the normal state, where the resulting self-consistent
equations are 
\begin{eqnarray}
\chi_m^\prime&=&T\sum_{n}\lambda_{m-n}\cr
&\int& d\epsilon
\frac{\rho(\epsilon)}{\rho(E_F)}\frac{\mu_N-\epsilon-\chi_n^\prime}{\omega_n^2
Z_n^{\prime 2}+
(\mu_N-\epsilon-\chi_n^\prime)^2},
\label{eq: chi_n_imag}
\end{eqnarray}
\begin{eqnarray}
Z_m^\prime&=&1+\frac{T}{\omega_m}\sum_{n}\lambda_{m-n}\cr
&\int& d\epsilon
\frac{\rho(\epsilon)}{\rho(E_F)}\frac{\omega_nZ_n^\prime}{\omega_n^2Z_n^{\prime 2}
+ (\mu_N-\epsilon-\chi_n^\prime)^2}.
\label{eq: z_n_imag}
\end{eqnarray}
A prime indicates the normal-state perturbation theory, 
$i\omega_n=i\pi T(2n+1)$ is the Fermionic
Matsubara frequency, $\chi_m^\prime=\chi^\prime(i\omega_m)=
{\rm Re}\Sigma^\prime(i\omega_m)$
is the real part of the normal-state self-energy, and $Z_m^\prime=Z^\prime
(i\omega_m)=1-{\rm Im} \Sigma^\prime(i\omega_m)/\omega_m$ 
is the so-called renormalization function determined
from the imaginary part of the normal-state self-energy. The symbols 
$\lambda_{m-n}$ are the electron-phonon coupling strengths
\begin{eqnarray}
\lambda_{m-n}&=&\lambda(i\omega_m-i\omega_n)\cr
&=&\int_0^\infty d\Omega
\alpha^2F(\Omega)\frac{2\Omega}{\Omega^2+(\omega_m-\omega_n)^2},
\label{eq: lambda_def}
\end{eqnarray}
and $\lambda=\lambda_0$.  The function $\rho(\epsilon)$ is
the electronic density of states as determined by the band-structure 
calculation,
and $\rho(E_F)$ is the density of states at the $T=0$ chemical potential of
the band-structure calculation ($E_F$). The chemical potential in the 
normal state is $\mu_N$.
In the limit as $T\rightarrow 0$, the number of Matsubara
frequencies becomes an infinite continuum, and the summations can be
replaced by integrals.  We find  
\begin{eqnarray}
\chi_0^\prime(i\omega)&=&\frac{1}{2\pi}\int d\omega^\prime
\lambda(i\omega-i\omega^\prime)\cr
&\int& d\epsilon
\frac{\rho(\epsilon)}{\rho(E_F)}\frac{\mu_{N0}-\epsilon-\chi_0^\prime
(i\omega^\prime)}{\omega^{\prime 2} Z_0^{\prime 2}(i\omega^\prime)+
[\mu_{N0}-\epsilon-\chi_0^\prime(i\omega^\prime)]^2},
\label{eq: chi_n_imag0}
\end{eqnarray}
\begin{eqnarray}
Z_0^\prime(i\omega)&=&1+\frac{1}{2\pi\omega}\int d\omega^\prime
\lambda(i\omega-i\omega^\prime)\cr
&\int& d\epsilon
\frac{\rho(\epsilon)}{\rho(E_F)}\frac{\omega^\prime
Z_0^\prime(i\omega^\prime)}{\omega^{\prime 2}
Z_0^{\prime 2}(i\omega^\prime)
+ [\mu_{N0}-\epsilon-\chi_0^\prime(i\omega^\prime)]^2},
\label{eq: z_n_imag0}
\end{eqnarray}
with the subscript zero denoting the results are at $T=0$. 
Evaluating these integrals with quadrature routines 
is much more efficient than calculating
the Matsubara sums with a large energy cutoff at finite temperature. 

Our strategy for determining the functions $\chi_0^\prime(i\omega)$ and 
$Z_0^\prime(i\omega)$
is to create a nonuniformly spaced grid on the imaginary axis (we use
336 points), with an upper
cutoff many times the electronic bandwidth. The grid is constructed in the 
following fashion.  The first grid point is chosen to lie at $i\omega_0=0$. 
Subsequent grid points are chosen by adding the new step size to the old
grid point $i\omega_{j+1}=i\omega_j+i\delta_j$, with $\delta_j$ increasing by
a factor of 1.1 with each step from its initial value $\delta_0=0.09$~meV
[hence $\delta_j=0.09\times
(1.1)^j$]. We restrict $\delta_j<400$~meV---if $\delta_j$
would be larger than 400~meV, we set it equal to 400~meV. The grid is reflected
about $i\omega=0$ to construct the negative frequency axis.
The integrals are then evaluated using a simple
Riemann sum over the nonuniform grid.  Once the functions $\chi_0^\prime$ and 
$Z_0^\prime$ are known
on the grid points, we linearly interpolate to evaluate them
at any point on the imaginary axis.  

We employ the $T=0$ normal-state solutions as an approximate solution for
high frequency.  This is included by first calculating
the normal-state perturbation theory at finite temperature, using the same 
Matsubara frequency cutoff as used in the superconducting state
$\omega_c=6\Omega_{\rm max}$, and then adding the difference between the 
$T=0$ normal state solution and the finite-$T$ solution to the superconducting
solution, as shown below.  We solve the following self-consistent 
equations for the normal-state self-energy at temperature $T$:
\begin{eqnarray}
\chi_m^\prime&=&T\sum_{|\omega_n|<\omega_c}\lambda_{m-n}
\int d\epsilon \frac{\rho(\epsilon)}{\rho(E_F)}\cr
&~&\Biggr [
\frac{\mu_N-\epsilon-\chi_n^\prime}{\omega_n^2 Z_n^{\prime 2}+
(\mu_N-\epsilon-\chi_n^\prime)^2}\cr
&~&-\frac{\mu_{N0}-\epsilon-\chi_0^\prime(i\omega_n)}{\omega_n^2 Z_0^{\prime 2}
(i\omega_n)+ [\mu_{N0}-\epsilon-\chi_0^\prime(i\omega_n)]^2}\Biggr ]\cr
&+&\chi_0^\prime(i\omega_m),
\label{eq: chi_n_imag_t}
\end{eqnarray}
\begin{eqnarray}
Z_m^\prime&=&1+\frac{T}{\omega_m}\sum_{|\omega_n|<\omega_c}\lambda_{m-n}
\int d\epsilon \frac{\rho(\epsilon)}{\rho(E_F)}\cr
&~&\Biggr [\frac{\omega_nZ_n^\prime}
{\omega_n^2Z_n^{\prime 2} + (\mu_N-\epsilon-\chi_n^\prime)^2}\cr
&~&-\frac{\omega_nZ_0^\prime(i\omega_n)}
{\omega_n^2Z_0^{\prime 2}(i\omega_n) + (\mu_{N0}-\epsilon-\chi_0^\prime)^2}
(i\omega_n)\Biggr ]\cr
&+&[Z_0^\prime(i\omega_m)-1].
\label{eq: z_n_imag_t}
\end{eqnarray} 
These equations would be exact if the summations over Matsubara frequencies
for the $T=0$ quantities were replaced by integrals.  Since the 
normal-state self-energy 
does not depend too strongly on $T$ for low temperature,
this approximation is accurate for low $T$. Note that the chemical
potential $\mu_N$ typically changes by about one meV (at $T=1.2$~K)
from the zero-temperature value $\mu_{N0}$ (in the nonconstant density of
states case).

The final set of equations we need are for the superconducting phase.
The self-consistent equations are calculated ``relative to the normal
state'' at $T$:
\begin{eqnarray}
\chi_m&=&T\sum_{|\omega_n|\le \omega_c}\lambda_{m-n}
\int d\epsilon \frac{\rho(\epsilon)}{\rho(E_F)}\cr
&~&\Biggr [ 
\frac{\mu_S-\epsilon-\chi_n}{\omega_n^2Z_n^2+
(\mu_S-\epsilon-\chi_n)^2+\Delta_n^2Z_n^2}\cr
&-&\frac{\mu_N-\epsilon-\chi_n^\prime}{\omega_n^2 Z_n^{\prime 2}+
[\mu_N-\epsilon-\chi_n^\prime]^2}\Biggr ]
+\chi_m^\prime,
\label{eq: chi_sc_imag}
\end{eqnarray}
\begin{eqnarray}
Z_m&=&1+\frac{T}{\omega_m}\sum_{|\omega_n|\le \omega_c}\lambda_{m-n}
\int d\epsilon \frac{\rho(\epsilon)}{\rho(E_F)}\cr
&~&\Biggr [\frac{\omega_nZ_n}{\omega_n^2Z_n^2+
(\mu_S-\epsilon-\chi_n)^2+\Delta_n^2Z_n^2}\cr
&-&\frac{\omega_nZ_n^\prime}{\omega_n^2Z_n^{\prime 2}
+ [\mu_N-\epsilon-\chi_n^\prime]^2}\Biggr ]
+[Z_n^\prime-1],
\label{eq: z_sc_imag}
\end{eqnarray}
and
\begin{eqnarray}
\Delta_mZ_m&=&T\sum_{|\omega_n|\le \omega_c}(\lambda_{m-n}-\mu^*)
\int d\epsilon \frac{\rho(\epsilon)}{\rho(E_F)}\cr
&~&\frac{\Delta_nZ_n}{\omega_n^2Z_n^2+
(\mu_S-\epsilon-\chi_n)^2+\Delta_n^2Z_n^2}.
\label{eq: delta_sc_imag}
\end{eqnarray}    
Here $\Delta_m=
\Delta(i\omega_m)=\Sigma_{12}(i\omega_m)/Z_m$ is the superconducting gap
determined from the off-diagonal self-energy, $\mu^*$ is the Coulomb
pseudopotential, and $\mu_S$ is the chemical potential 
in the superconducting state. Note that we add and subtract
the normal-state results at finite temperature
in order to ensure that the Matsubara frequency 
summations converge rapidly ($\omega_c=6\Omega_{\rm max}$ is the cutoff 
frequency; the high-frequency tails are already included in the
normal-state self-energy). This procedure allows for a rapid computation of 
the many-body Green's functions when there is an energy-dependent 
electronic density of states.
We only calculate the Green's functions at the Matsubara frequencies here.

The chemical potentials in the normal and superconducting states
are determined by the requirement
that the electron density be equal to the equilibrium density of electrons for
the given band.  Our first step is to find the $T=0$ Fermi level
for the band-structure density of states, which satisfies
\begin{equation}
\rho_e=2\int_{-\infty}^{E_F}d\epsilon \rho(\epsilon),
\label{eq: band_ne}
\end{equation}
with the factor of 2 coming  from spin.  
In the normal state at $T=0$, we use the
$T\rightarrow 0$ limit of the
identity $\rho_e=2T\sum_nG(i\omega_n)$ and Eq.~(\ref{eq: band_ne}) to
produce the self-consistent equation for $\mu_{N0}$:
\begin{eqnarray}
2\int_{\mu_{N0}}^{E_F}d\epsilon\rho(\epsilon)&=&
\frac{1}{\pi}\int d\epsilon\rho(\epsilon)\int d\omega\cr
&~&\Biggr [ \frac{1}{i\omega Z_0^\prime(i\omega)+\mu_{N0}
-\epsilon+\chi_0^\prime
(i\omega)}\cr
&~&-\frac{1}{i\omega+\mu_{N0}-\epsilon}\Biggr ].
\label{eq: normal_ne}
\end{eqnarray}
The normal-state chemical potential
at finite $T$ (we use $T=1.2$~K for the tunneling
inversion) is found by comparison with the
normal state at $T=0$: $0=2T\sum_n[G(i\omega_n)-G(i\omega_n)|_{T=0}]$ (note the
$T=0$ sum is an approximation to the continuum integral).
Finally, for the superconducting state, we use the comparison of the
normal-state filling with the superconducting-state filling to find
$\mu_S$:
\begin{eqnarray}
0&=&\frac{T}{\pi}\int d\epsilon\rho(\epsilon)\sum_{|\omega_m|<\omega_c}\cr
&~&\Biggr [ \frac{\mu_S-\epsilon-\chi_m}{\omega_m^2Z_m^2+(\mu_S-\epsilon-
\chi_m)^2+\Delta_m^2Z_m^2}\cr
&~&-\frac{\mu_N-\epsilon+\chi_m^\prime}{\omega_m^2 Z_m^{\prime 2}+
(\mu_N-\epsilon+ \chi_m^\prime)^2} \Biggr ] .
\label{eq: sc_ne}
\end{eqnarray}

The next step is to calculate the self-energy on the real axis using
an exact analytic continuation technique.\cite{marsiglio}  We begin with the
normal state at $T=0$.  The self-energy satisfies
\begin{eqnarray}
\Sigma_0^\prime(\omega&+&i\eta)=\frac{1}{2\pi\rho(E_F)}\int_{-\infty}^{\infty}
d\omega^\prime\lambda(\omega-i\omega^\prime)\cr
&\times&\int d\epsilon
\frac{\rho(\epsilon)}{i\omega^\prime Z_0^\prime(i\omega^\prime)+
\mu_{N0}-\epsilon-\chi_0^\prime(i\omega^\prime)}\cr
&+&\frac{1}{\rho(E_F)}\int_0^\omega d\Omega\alpha^2F(\Omega)
\int d\epsilon\cr
&\times&
\frac{\rho(\epsilon)}{(\omega-\Omega)Z_0^\prime(\omega-\Omega)+\mu_{N0}-\epsilon
-\chi_0^\prime(\omega-\Omega)+i\eta},
\label{eq: normal_real_sigma}
\end{eqnarray}
where $\eta\rightarrow 0^+$.
This is a self-consistent equation, because the second integral contains
the self-energy on the real axis from the definitions
\begin{equation}
Z_0^\prime(\omega)=1-\frac{\Sigma_0^\prime(\omega+i\eta)-
\Sigma_0^{\prime*}(-\omega+i\eta)}{2\omega},
\label{eq: z_norm_real}
\end{equation}
and
\begin{equation}
\chi_0^\prime(\omega)=\frac{\Sigma_0^\prime(\omega+i\eta)+
\Sigma_0^{\prime*}(-\omega+i\eta)}{2},
\label{eq: chi_norm_real}
\end{equation}
where the $*$ denotes complex conjugation.  The term $\lambda(\omega-
i\omega^\prime)$ is found from the spectral formula and the given
Eliashberg function $\alpha^2F(\Omega)$:
\begin{equation}
\lambda(\omega-i\omega^\prime)=-\int_{-\infty}^\infty d\Omega\frac{\alpha^2F
(\Omega)}{\omega-i\omega^\prime-\Omega}.
\label{eq: lambda_spectral}
\end{equation}

Now we calculate the finite-$T$ normal-state results relative to
the $T=0$ calculation:
\begin{eqnarray}
\Sigma^\prime(\omega&+&i\eta)=\frac{T}{\rho(E_F)}\sum_{|\omega_n|<\omega_c}
\lambda(\omega-i\omega_n)\cr
&\times&\int d\epsilon\Biggr [
\frac{\rho(\epsilon)}{i\omega_n Z_n^\prime+
\mu_N-\epsilon-\chi_n^\prime}\cr
&~&-\frac{\rho(\epsilon)}{i\omega_n Z_0^\prime
(i\omega_n)+ \mu_{N0}-\epsilon-\chi_0^\prime(i\omega_n)}\Biggr ]\cr
&+&\frac{1}{2\rho(E_F)}\int_{-\infty}^\infty d\Omega\alpha^2F(\Omega)
\cr
&~& ~\left [ {\rm tanh} \left ( \frac{\omega-\Omega}{2T} \right )+
{\rm coth} \left ( \frac{\Omega}{2T} \right ) \right ]
\int d\epsilon\rho(\epsilon)\cr
&\times&\Biggr [
\frac{1}{(\omega-\Omega)Z^\prime(\omega-\Omega)+\mu_{N}-\epsilon
-\chi^\prime(\omega-\Omega)+i\eta}\cr
&-&
\frac{1}{(\omega-\Omega)Z_0^\prime(\omega-\Omega)+\mu_{N0}-\epsilon
-\chi_0^\prime(\omega-\Omega)+i\eta}\Biggr ]\cr
&+&\Sigma_0^\prime(\omega+i\eta),
\label{eq: normal_real_sigma_t}
\end{eqnarray}
where the $Z$ and $\chi$ functions on the real axis are determined from 
equations analogous to Eqs.~(\ref{eq: z_norm_real}) and 
(\ref{eq: chi_norm_real}).

For the superconducting state,
calculations are performed 
relative to the normal state at $T$:
\begin{eqnarray}
&~&\Sigma_{11}(\omega+i\eta)=
T\sum_{|\omega_n|<\omega_c}\lambda(\omega-i\omega_n) \cr
&\times&\int d\epsilon\rho(\epsilon)\Biggr [
\frac{i\omega_n Z_n+\mu_S-\chi_n}{\omega_n^2 Z_n^2+
(\mu_S-\epsilon-\chi_n)^2+\Delta_n^2Z_n^2}\cr
&~&-\frac{i\omega_n Z_n^\prime+\mu_N-\chi_n^\prime}{\omega_n^2 Z_n^{\prime 2}
+ (\mu_{N}-\epsilon-\chi_n^\prime)^2}\Biggr ]\cr
&+&\frac{1}{2\rho(E_F)}\int_{-\infty}^\infty d\Omega\alpha^2F(\Omega)
\cr
&~& ~\left [ {\rm tanh} \left ( \frac{\omega-\Omega}{2T} \right )+
{\rm coth} \left ( \frac{\omega-\Omega}{2T} \right ) \right ]
\int d\epsilon\rho(\epsilon)\cr
&\Biggr [&
\frac{(\omega-\Omega)Z(\omega-\Omega)+\mu_{S}-\epsilon
-\chi(\omega-\Omega)}{(\omega-\Omega)^2Z^2(\omega-\Omega)+[\mu_{S}-\epsilon
-\chi(\omega-\Omega)]^2+\Delta(\omega-\Omega)^2Z(\omega-\Omega)^2}\cr
&-&
\frac{(\omega-\Omega)Z^\prime(\omega-\Omega)+\mu_{N}-\epsilon
-\chi^\prime(\omega-\Omega)}{(\omega-\Omega)^2Z^{\prime 2}
(\omega-\Omega)+[\mu_{N}-\epsilon
-\chi^\prime(\omega-\Omega)]^2}\Biggr ]\cr
&+&\Sigma^\prime(\omega+i\eta),
\label{eq: sc_11_real}
\end{eqnarray}
for the diagonal self-energy and
\begin{eqnarray}
&~&\Sigma_{12}(\omega+i\eta)=T\sum_{|\omega_n|<\omega_c}[
\lambda(\omega-i\omega_n)-\mu^*]
\cr
&\times&\int d\epsilon\rho(\epsilon)
\frac{\Delta_nZ_n}{\omega_n^2 Z_n^2+
(\mu_S-\epsilon-\chi_n)^2+\Delta_n^2Z_n^2}\cr
&+&\frac{1}{2\rho(E_F)}\int_{-\infty}^\infty d\Omega\alpha^2F(\Omega)
\cr
&~& ~\left [ {\rm tanh} \left ( \frac{\omega-\Omega}{2T} \right )+
{\rm coth} \left ( \frac{\omega-\Omega}{2T} \right ) \right ]
\int d\epsilon\rho(\epsilon)\cr
&\times&
{\Delta(\omega-\Omega)Z(\omega-\Omega)}\cr
&/&
\{(\omega-\Omega)^2Z^2(\omega-\Omega)+[\mu_{S}-\epsilon
-\chi(\omega-\Omega)]^2\cr
&~&+\Delta(\omega-\Omega)^2Z(\omega-\Omega)^2\},
\label{eq: sc_12_real}
\end{eqnarray}
for the off-diagonal self-energy.

The tunneling conductance satisfies
\begin{equation}
\frac{[dI/dV]_S}{[dI/dV]_N}(\omega)={\rm Re} \left [
\frac{\omega}{\sqrt{\omega^2- \Delta^2(\omega)}}\right ].
\label{eq: conductance}
\end{equation}
The reduced density of states (RDOS) is the ratio of the tunneling conductance
to the BCS tunneling conductance minus one, which becomes
\begin{equation}
RDOS(\omega)={\rm Re}\left [
\frac{\sqrt{\omega^2-\Delta_0^2}}{\sqrt{\omega^2-\Delta^2(\omega)}}\right ]
-1,
\label{eq: rdos}
\end{equation}
where $\Delta_0$ is the superconducting gap, defined as the solution to the
equation ${\rm Re}\Delta(\omega)=\omega$ (at $\omega=\Delta_0$).

\begin{figure}[htbf]
\epsfxsize=2.5in
\centerline{\epsffile{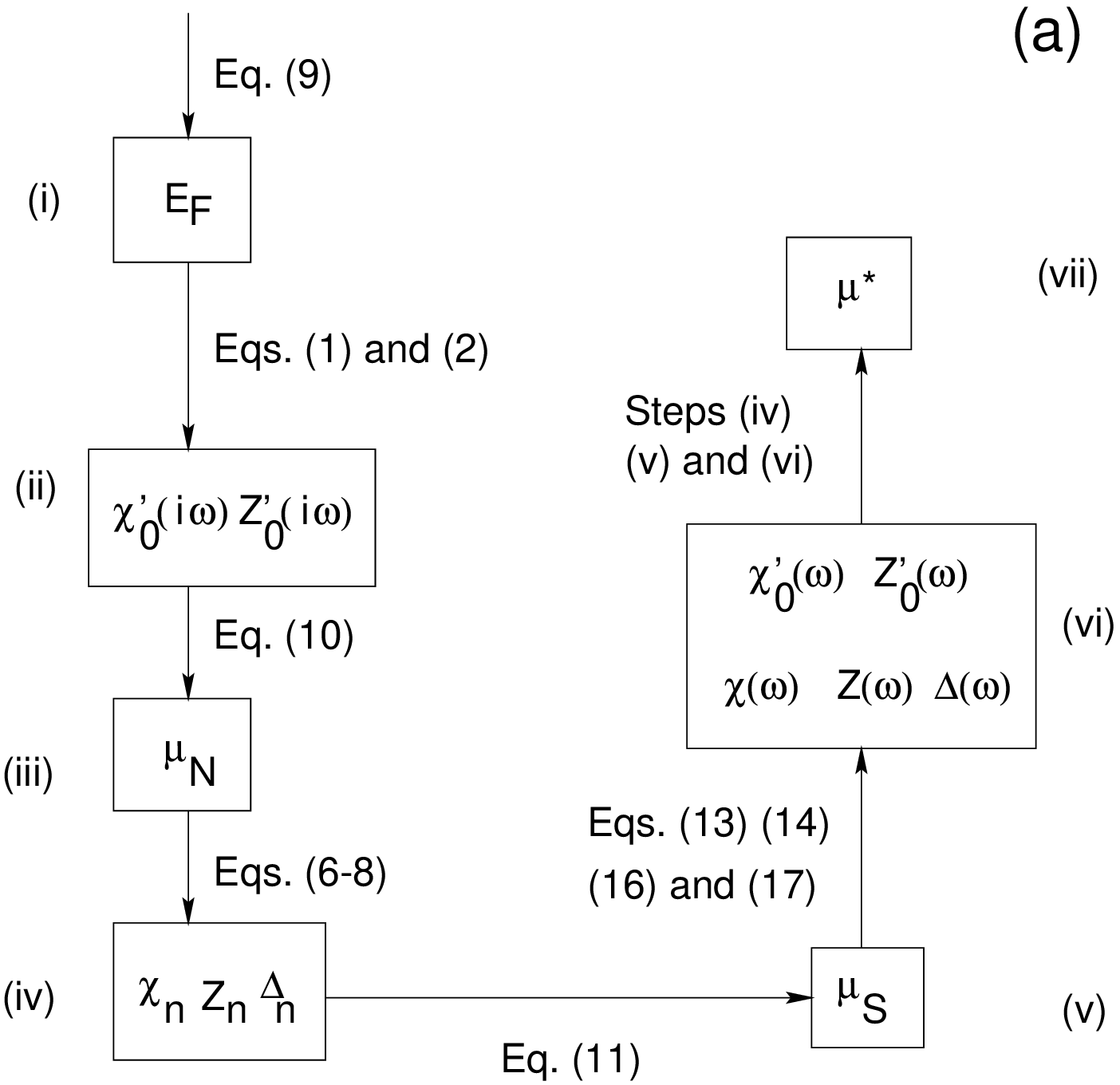}}
\epsfxsize=3.3in
\centerline{\epsffile{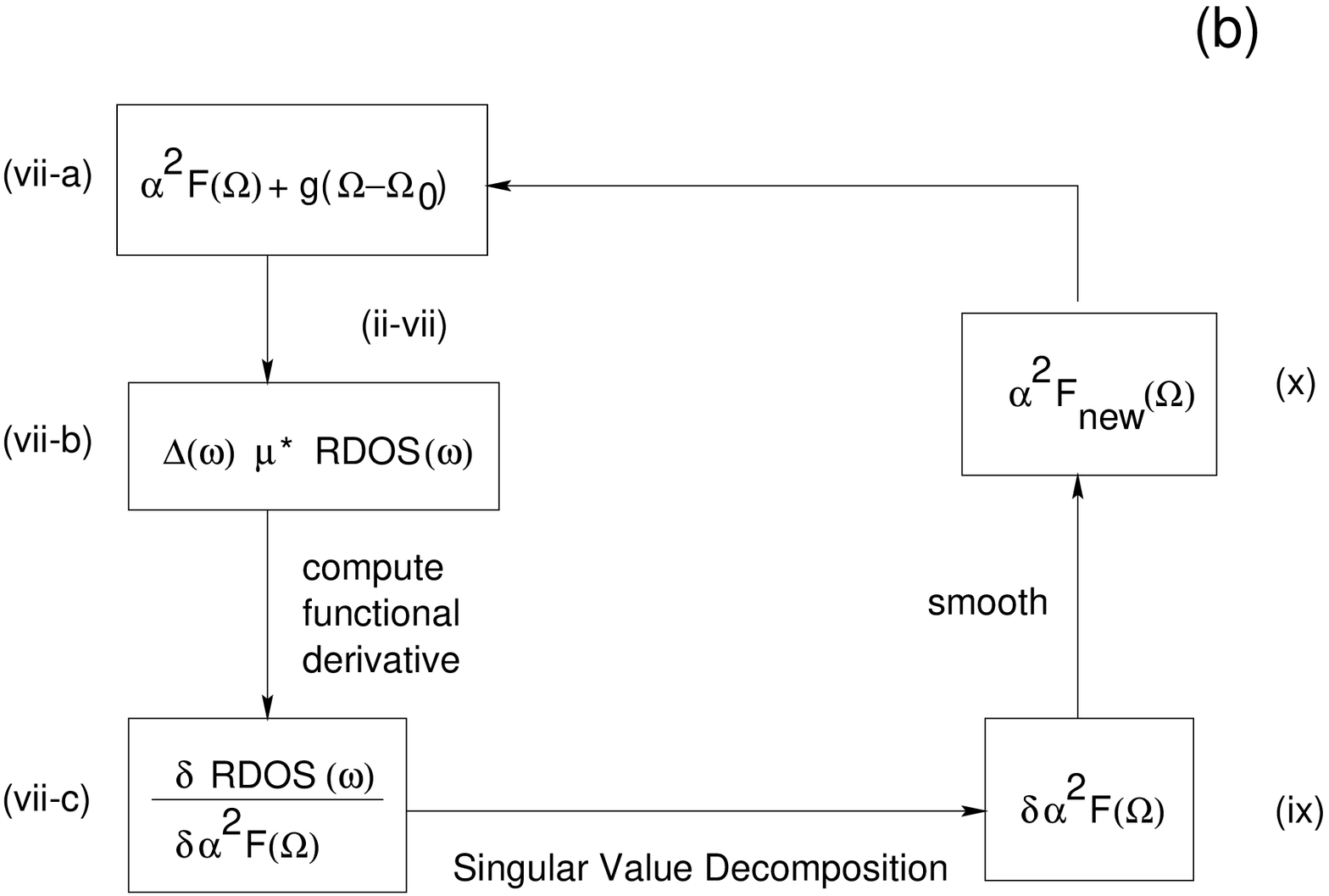}}
\vskip 0.10in
\caption{
Block diagrams of the algorithm employed in the modified McMillan-Rowell
tunneling inversion with a nonconstant electronic density of states. In panel
(a), we show the main algorithm used to determine the RDOS.  Steps (i-vii)
are described in detail in the text.  Panel (b) shows the McMillan-Rowell
strategy for updating the Eliashberg function in the tunneling inversion, with
details described in the text.
}
\label{fig: algorithm}
\end{figure}

To solve all of these self-consistent
equations and to invert the tunneling conductance to get the 
Eliashberg function, we
begin with the LDA band-structure electronic density of states as input.
We take the band to consist of the states that contain nine electrons
below the Fermi level and four electrons above, corresponding
to approximately $\pm 0.8$~eV about $E_F$.  Hence the 
electron filling
satisfies $\rho_e=9$.  Our first step (i) is to determine the band-structure
Fermi level $E_F$ at $T=0$ by solving Eq.~(\ref{eq: band_ne}).  
Then we take two initial
guesses for $\mu_N$.  (ii) For each $\mu_N$, we self-consistently solve
for Eqs.~(\ref{eq: chi_n_imag}) and (\ref{eq: z_n_imag}) at $T=0$.  Once those
functions are known, (iii) we use a one-dimensional root-finder to solve 
Eq.~(\ref{eq: normal_ne}) to determine $\mu_N$. The next step is to determine 
the superconducting chemical potential $\mu_S$ and the Coulomb pseudopotential
$\mu^*$.  We first choose a guess for $\mu^*$.  With $\mu^*$ fixed, we choose
two initial guesses for $\mu_S$ and (iv) self-consistently
solve Eqs.~(\ref{eq: chi_sc_imag}), (\ref{eq: z_sc_imag}), and 
(\ref{eq: delta_sc_imag}).  Once those functions are determined, (v) we use
a one-dimensional root-finder to solve Eq.~(\ref{eq: sc_ne}) and find $\mu_S$.
(vi) Now the real-axis equations for both the normal state at $T=0$,
Eqs.~(\ref{eq: z_norm_real}) and (\ref{eq: chi_norm_real}), and the normal and
superconducting states at $T$,
Eqs.(\ref{eq: normal_real_sigma_t}), (\ref{eq: sc_11_real}), and 
(\ref{eq: sc_12_real}), are solved self-consistently.  Once the gap
function $\Delta(\omega)$ is known, we can find the superconducting gap
$\Delta_0$. (vii) Steps (iv), (v), and (vi) are repeated for different values of 
$\mu^*$ until we find the particular value where $\Delta_0=3.15$~meV, which
is the superconducting gap for Nb$_3$Sn at low temperature.  
If we knew the correct Eliashberg
function, then steps (i)-(vii) would be all that we need to calculate the
RDOS from Eq.~(\ref{eq: rdos}).  But we need to find the best fit 
$\alpha^2F(\Omega)$ that is consistent with the experimental RDOS.  To do
this, we simply employ the McMillan-Rowell procedure for the tunneling
inversion.  We start with a guess for $\alpha^2F$.  (viii) Next we
calculate the functional derivative of how changes in the Eliashberg function
affect the RDOS.  The functional derivative is found by  adding
a small-weight Gaussian to $\alpha^2F(\Omega)$ centered at a given frequency
$\Omega_0$,  repeating steps (ii-vii) to determine the RDOS,
and  calculating the functional derivative by taking the difference 
of the new RDOS with the RDOS for the original Eliashberg function. 
This step (viii) is repeated for each frequency in the
discrete $\alpha^2F$ to determine the functional derivative matrix.  The next
step (ix) is to choose the weights for adjusting $\alpha^2F$ to fit the
experimental RDOS better.  Since the functional derivative matrix may be
rank-deficient, we use a singular-value decomposition to determine the
updated weights.  Once the updated $\alpha^2F(\Omega)$ is found, (x) we apply
an exploratory-data-analysis robust smoother followed by a Hanning smoother.
Steps (viii-x)
are repeated until the updated $\alpha^2F(\Omega)$ ceases to produce a RDOS
that is closer to experiment.  We force the Eliashberg function to 
be positive everywhere and to be
quadratic for $\Omega<2.5$~meV. 
The step size for the discrete frequencies
at which  $\alpha^2F$ is evaluated is 0.45~meV.  The algorithm is depicted
pictorially in Fig.~\ref{fig: algorithm}.

There is a numerical problem with this algorithm.  In general, the 
weights for the shift in $\alpha^2F$ are strongly oscillatory, which 
forces the Eliashberg function to have large-amplitude narrow-width peaks.  
Since we do not expect the Eliashberg function to have such sharp peaks, they
need to be smoothed away in the next update.  Because the functional derivatives
are not calculated with this smoothing procedure (and there is no obvious
way to incorporate
the smoothing into the derivatives), we are limited in how closely we can
reproduce  
the experimental RDOS,  even though we have the full freedom to adjust
$\alpha^2F$.  

\section{Results}

\begin{figure}[htbf]
\epsfxsize=3.5in
\centerline{\epsffile{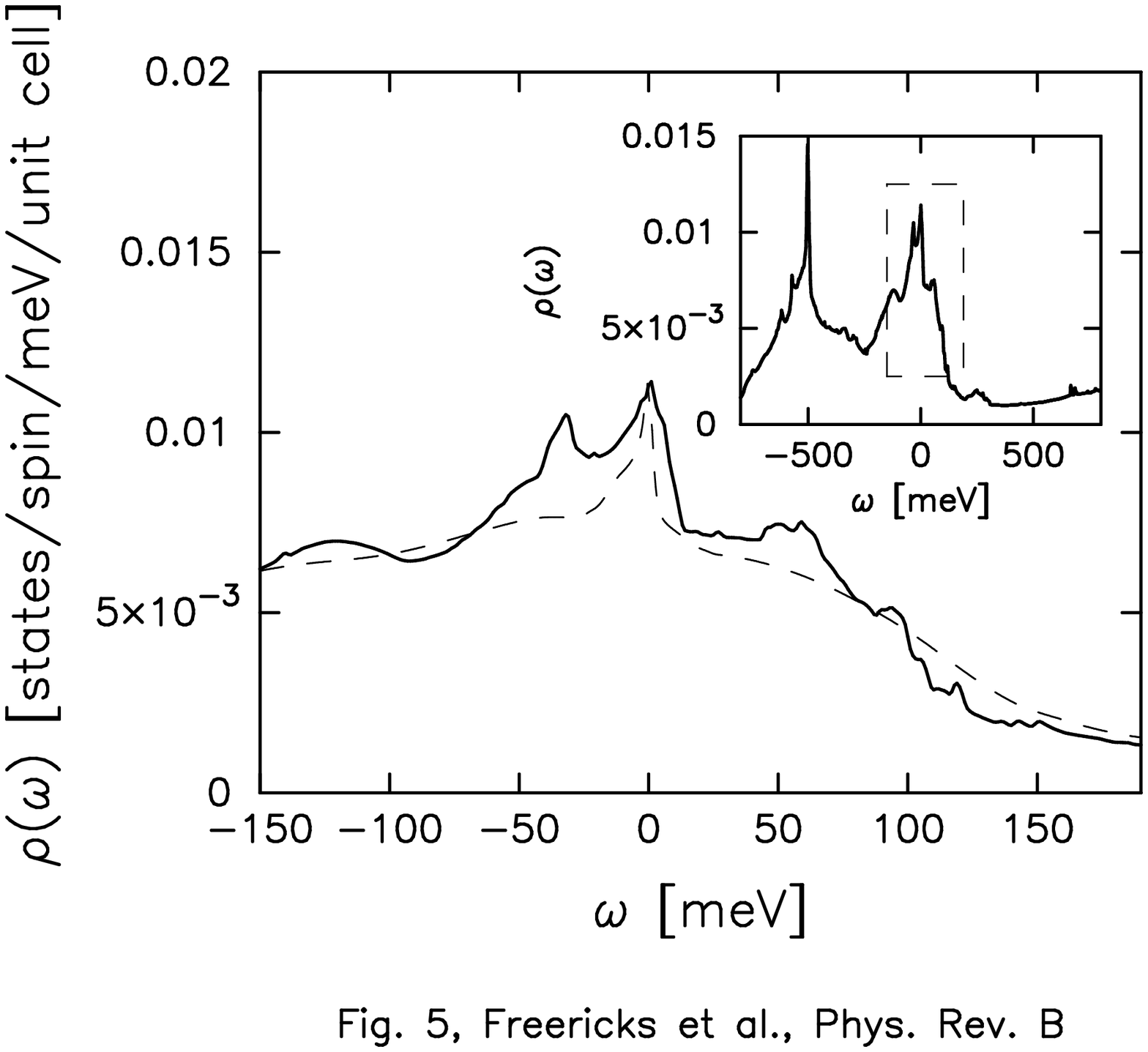}}
\caption{
Density of states for Nb$_3$Sn.  The solid curve is the band-structure
density of states as calculated with density functional theory.  The
dashed curve is the quasiparticle density of states calculated with
the fit $\alpha^2F$.  The chemical potential lies at $\omega=0$.
Inset shows the full band-structure
density of states, with the dashed box indicating the region blown up in the main figure.
The units of the vertical axis for both the main plot and the inset are 
states per spin per meV per unit cell.
}
\label{fig: dos}
\end{figure}

We first consider a test case, where the experimental
tunneling data for lead is used to perform both a conventional and a nonconstant
DOS tunneling inversion.  In both cases we use the same code
to perform the analysis as summarized in Fig.~\ref{fig: algorithm}.  In
the constant DOS case, we use the DOS for lead
($2.5\times 10^{-4}$ states per spin per meV per unit cell) and choose a band 
that contains
two total electrons (including spin) and $\rho_e=1$.  In the nonconstant 
DOS 
case, we use the electronic DOS for Nb$_3$Sn (see inset
to Fig.~\ref{fig: dos}).  This is of course   
an artificial problem that is presented for illustrative purposes.

\begin{figure}[htbf]
\epsfxsize=3.5in
\centerline{\epsffile{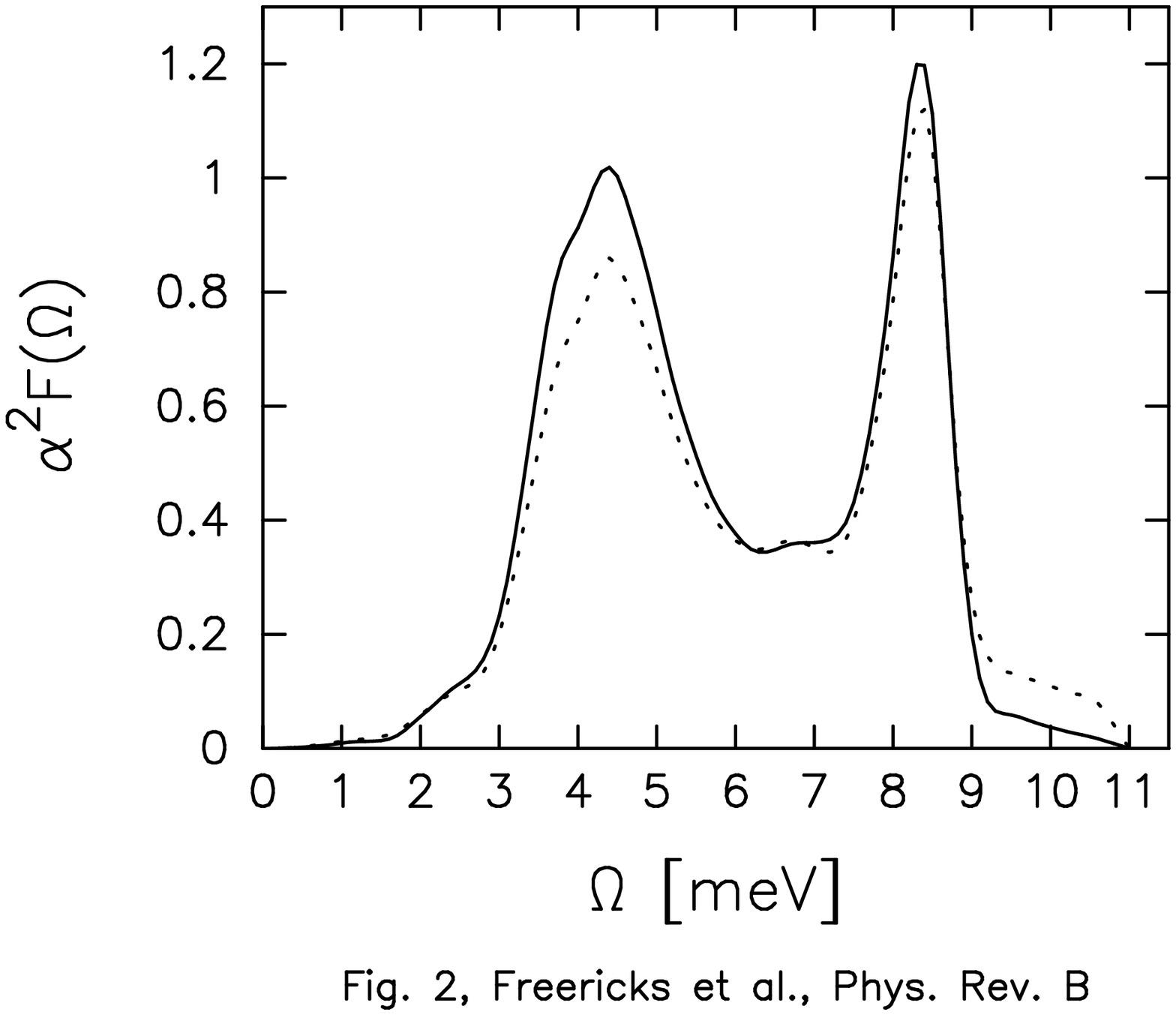}}
\caption{
Eliashberg functions for a tunneling inversion in lead with
(solid line) constant density of states and (dotted lines) the
nonconstant density of states of Nb$_3$Sn.  The general
shapes of the two curves are similar.
The values of
$\lambda$ are 1.56 and 1.42 for the constant and nonconstant  density of 
states cases, respectively. 
}
\label{fig: a2f_pb}
\end{figure}

The results for the best fit Eliashberg functions are given in 
Fig.~\ref{fig: a2f_pb}.  The solid line results from a tunneling inversion
done with a constant DOS, and is consistent with 
earlier analyses of the lead tunneling data
($\lambda=1.56$, maximal error of 0.002, and a root-mean-square error
of 0.0004).  The tunneling-inversion result
based on the energy-dependent Nb$_3$Sn DOS is plotted
as a dotted line, and corresponds to 
$\lambda=1.42$, with a maximal error of 0.004, and a
root-mean-square error of 0.0015.  The fit with a nonconstant DOS 
is  about four times worse than with the constant
DOS.  Note further, that the main effect of the sharp
peak in the DOS is to reduce the overall scale of
$\alpha^2F$ for nearly all frequencies except the highest, 
where it is strongly enhanced.  Qualitatively, the two
$\alpha^2F$ curves are very similar in shape.

The quality of the tunneling data for lead is so good that one can
actually see that the tunneling inversion works better with a constant
DOS than with a nonconstant DOS.  What is 
interesting is that one can get reasonable results using the wrong
nonconstant DOS due to the freedom allowed in choosing the 
function $\alpha^2F$.

\begin{figure}[htbf]
\epsfxsize=3.5in
\centerline{\epsffile{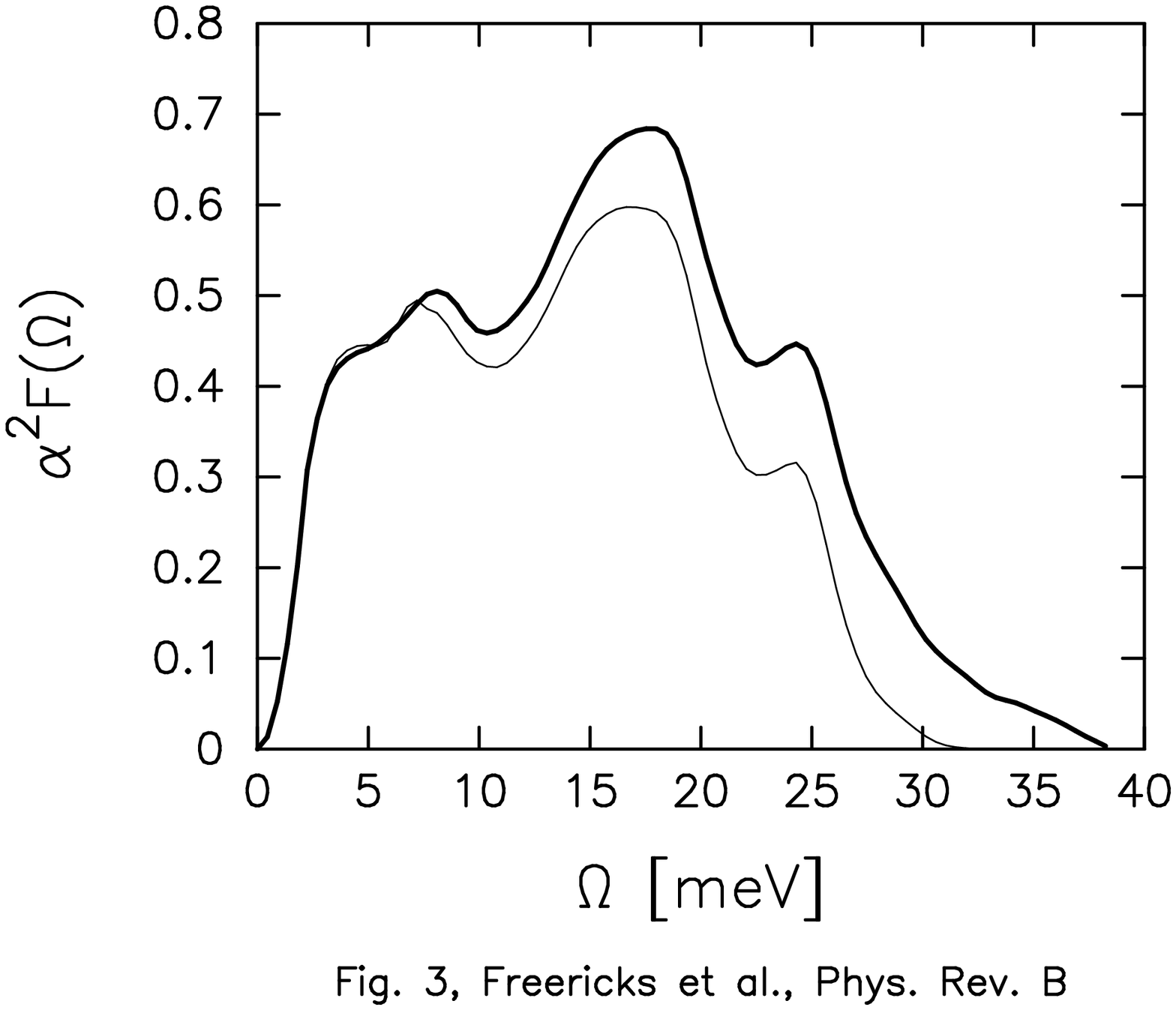}}
\caption{
Eliashberg functions for a tunneling inversion in Nb$_3$Sn for
nonconstant density of states (thick) and  
constant density of states (thin). 
In the nonconstant density  of states analysis, the spectral weight 
is enhanced at high energies.
}
\label{fig: a2f_a15}
\end{figure}

We next turn to the results for the tunneling inversion of Nb$_3$Sn.
The extracted Eliashberg functions are shown in Fig.~\ref{fig: a2f_a15}.
The thick curve shows the nonconstant DOS results, while the
thin curve is for the constant DOS case.
The properties of these electron-phonon
spectral functions are summarized in Table I.  
In contrast to the results for lead, 
where the Eliashberg function is reduced in the 
nonconstant DOS analysis, it is enhanced for Nb$_3$Sn.  The
main difference, aside from an overall scale factor, is a large enhancement
for $\Omega>20$~meV, particularly in the highest energies,
where the constant DOS $\alpha^2F$ is strongly suppressed. Note that the maximal
allowed phonon frequency is the DFT-calculated 38~meV in both cases; the fitting
procedure sharply suppresses the $\alpha^2F$ at high energy for the constant
DOS calculation. The
general shapes of these curves are similar to those found previously\cite{geerk}
but the overall scale is larger. The enhancement occurs because we are unable
to reproduce the full overswing.  In order to compensate for this, $\lambda$
is increased and the quality of the low-energy fit is reduced.  Note that we fit
all experimental points, including those in the overswing region 
($\omega>35$~meV)---nevertheless,
we are unable to get ``good'' results for the extracted $\alpha^2F$ in the sense
that the best fit $\alpha^2F$ is unable to accurately reproduce the tunneling
data over the entire experimentally measured range (due to the inability to
produce the full overswing).  

There are, nevertheless,  a number of promising features of this
calculation. If we compare Fig.~\ref{fig: a2f_a15}
to the data for low-$T_c$ (disordered) Nb$_3$Sn
junctions, we find the broad low-energy peak and shoulder (present from 4 to
8~meV) lies a few meV lower in the high-$T_c$ material than the low-$T_c$
one\cite{geerk} (where it lies at 10~meV).  This is the expected phonon
softening that leads to the high $T_c$ of the A15 compounds that is evident also
in second-derivative data\cite{second_deriv} on Nb$_3$Sn, where the low-$T_c$
material has a large peak at 10~meV which disperse to two peaks (one at 6 and 
one at 8~meV) in the high-$T_c$ material. Another strong indication of enhanced
electron-phonon coupling is the pronounced softening of the 10~meV shoulder
observed in the phonon density of states of Nb$_3$Sn on cooling down from
room temperature to 4.2~K as described by Schweiss et al.\cite{neutron}
Hence the extracted $\alpha^2F$
displays the expected phonon-mode softening, with the expected energy scales.
Another test is to compare the extracted $\alpha^2F$ to the phonon DOS, 
$F(\Omega)$, measured with neutron scattering\cite{neutron}. In actuality,
neutron scattering weights the phonons by the neutron scattering cross
section for each nucleus forming the generalized DOS $G(\Omega)$. Nb has a
cross section approximately twice as big as Sn, but as the atomic masses of
Nb and Sn are similar, we expect both atoms to be in motion for most phonon
modes. Hence these modes can be excited by interactions of neutrons with either
Nb or Sn nuclei.  This averaging effect implies that the phonon density of
states $F(\Omega)$ and the generalized density of states $G(\Omega)$ should
agree closely\cite{neutron,weber} for Nb$_3$Sn. If we take the ratio of the
tunneling data to the neutron data, we find a large peak in $\alpha^2(\Omega)$
for energies below 10~meV.  This agrees with theoretical
calculations\cite{weber}, which predict a large $\alpha^2(\Omega)$ for the 
low-energy phonon modes (below 10~meV).

\begin{figure}[htbf]
\epsfxsize=3.5in
\centerline{\epsffile{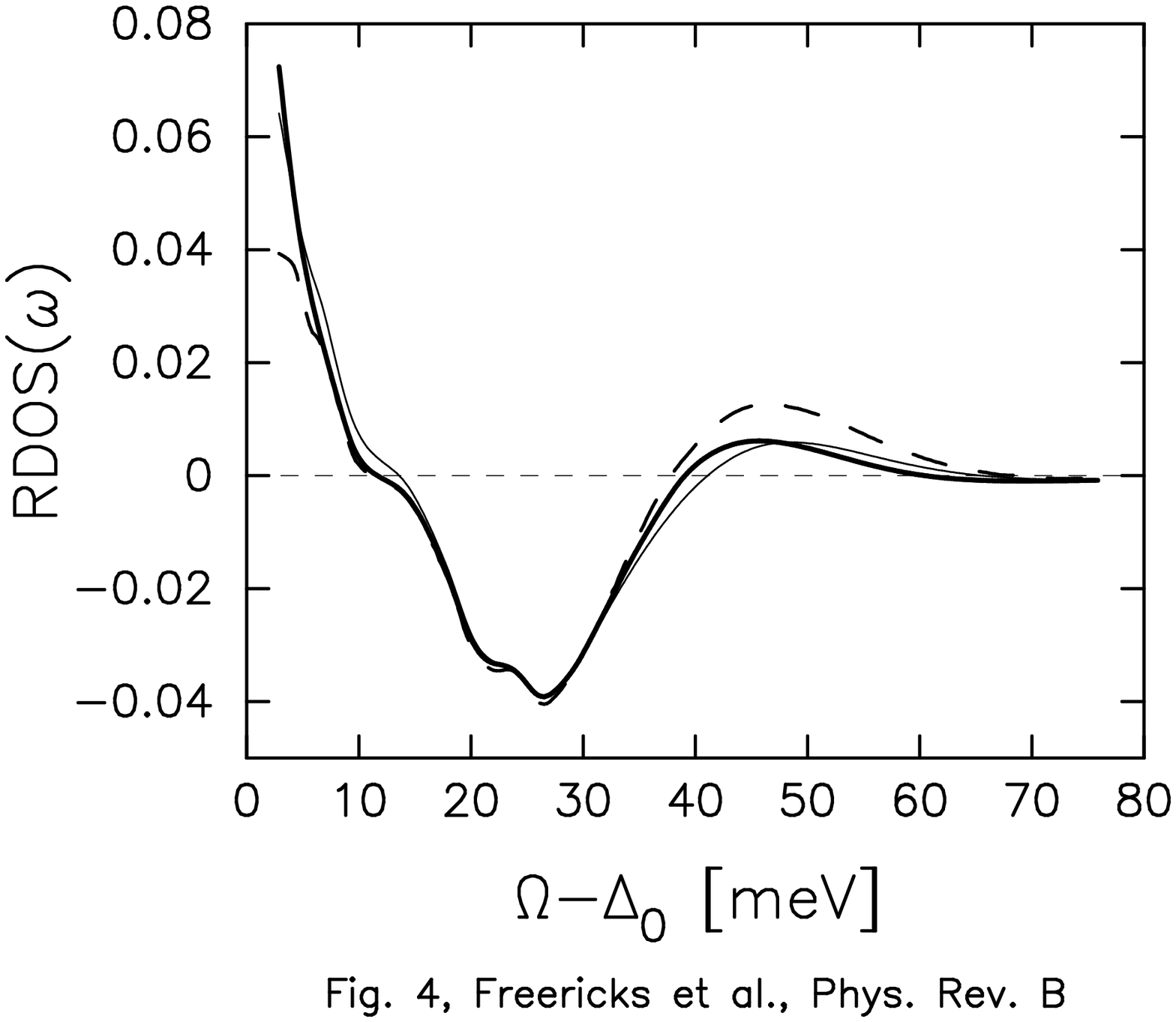}}
\caption{
Reduced density of states [from Eq.~(\ref{eq: rdos})] for  
nonconstant (thick) and constant (thin) density of states.
The thick dashed line is the experimental data (scanned and digitized) and
the thin dashed line is the horizontal axis.  Note the
poor quality of the fit at the lowest energies, and the sharper overswing
at high energies for the nonconstant density of states case. The poor fit arises
from the fact that $\lambda$ must be increased in order to
produce the overswing, but this reduces the agreement at low 
frequencies.  The results shown represent the best compromise for fitting the
entire experimental spectrum.
}
\label{fig: rdos}
\end{figure}

Next we consider the reduced density of states for both cases in 
Fig.~\ref{fig: rdos}.  The thick dashed curve is the
experimental data,\cite{geerk} the thin dashed line is the horizontal
axis, the thick solid line is the nonconstant DOS case and the
thin solid line is the constant DOS results. 
There are three important points to note 
about these curves; (i) the fits are poor at the lowest
energies, (ii) the nonconstant density of states produces a more
rapid ``overswing'' and return-to-zero at about 40~meV, and (iii) the amplitude 
of the overswing is enhanced in the region from $40-60$~meV.  As neither of the
fits is particularly good, we cannot conclude from this work that the
inclusion of 
nonconstant density of states effects alone is sufficient to 
explain the A15 tunneling data.  But we do see that the 
incorporation of a nonconstant density of states definitely provides 
improvements.  It is not clear what else is needed. Part of the problem may be 
with the numerical instabilities of the tunneling-inversion algorithm.
Alternatively, there may be an intrinsic thin proximity layer that always needs
to be taken into account regardless of the quality of the junction.
Other factors that may be important include anharmonicity, anisotropy, 
and impurity scattering.  It appears unlikely that disorder is the
explanation, since disorder tends to reduce the magnitude of the RDOS,
not increase it, as is needed.  The proximity-effect explanation is also 
hard to support, because the constant DOS analysis found the proximity
layer to be vanishingly thin.  If that conclusion holds true
for the nonconstant DOS
analysis as well, then this would not be a viable explanation either.
The effects of anharmonicity at low temperature should be explainable
within a quasiharmonic approximation, unless there is a preformed-pair
phase.  Hence, we believe the
most likely cause of the discrepancy is from anisotropic effects.  It is 
conceivable that tunneling barriers grow differently on different indexed
surfaces.  The strong dependence of the tunneling conductance on barrier thickness
thus could lead to a directional selectivity of the tunneling current and
thereby anisotropy effects would influence the tunneling current even in the
case of polycrystalline films.
Incorporation of such anisotropic effects is beyond the scope of this work.

We conclude with a discussion of the 
properties of the solutions to the many-body problem.  
The electronic density of states is shown in Fig.~\ref{fig: dos}.  The
solid curve is the band-structure density of states as calculated within
density-functional theory.  The dashed curve is the quasiparticle density
of states in the normal state at $T=0$ calculated with the fit $\alpha^2F$.
It depicts the behavior expected.
The density of states is unchanged at the chemical potential ($\omega=0$)
because the self-energy is momentum independent,
the peak is narrowed by a factor of about 3 (due to ``1+$\lambda$''
narrowing), and the overall density
of states is smoothed out due to lifetime effects.  In the inset, we show
the full band-structure density of states used in the calculations.  
The dashed box
indicates the region shown in the main figure.

\begin{figure}[htbf]
\epsfxsize=3.5in
\centerline{\epsffile{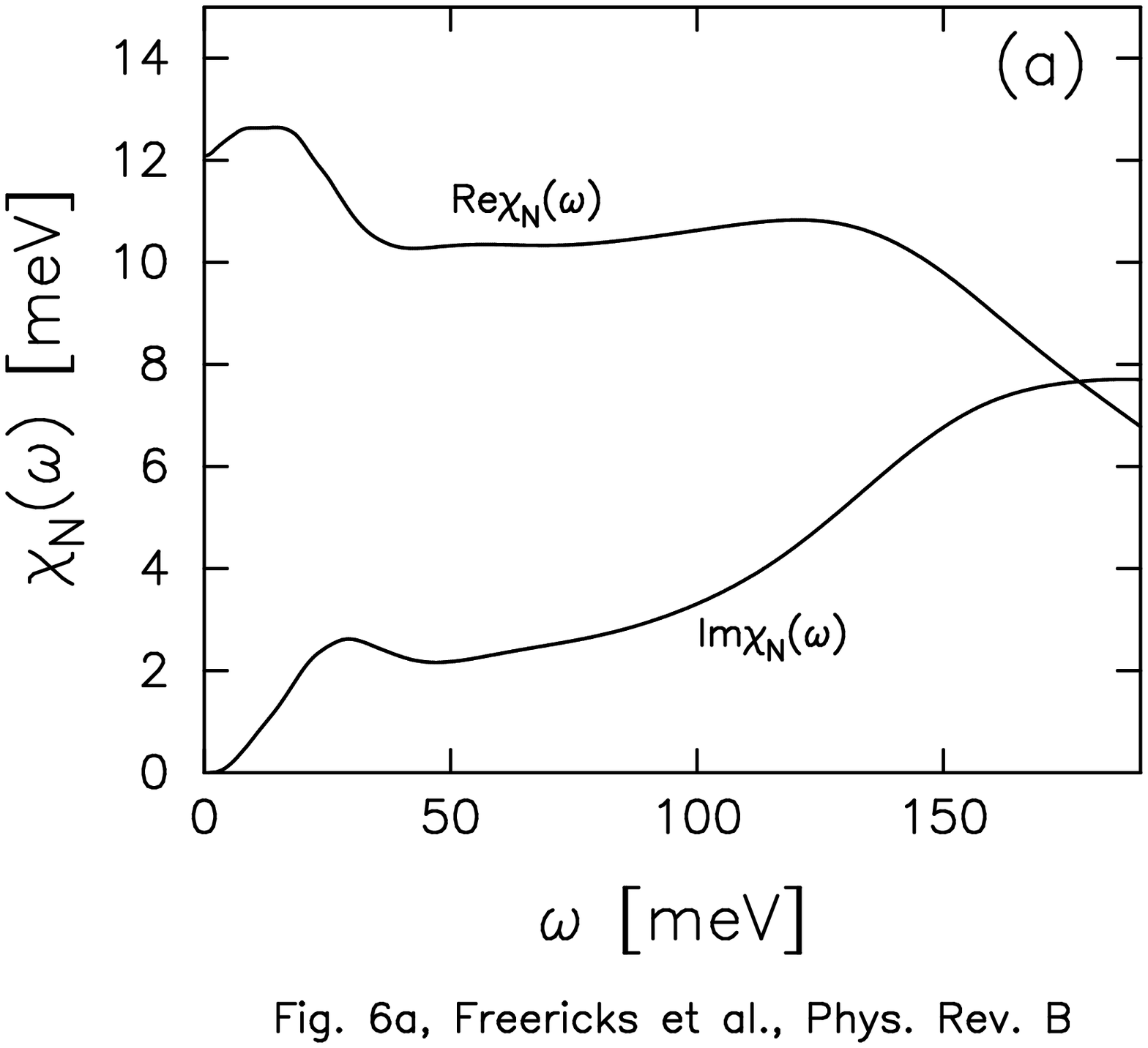}}
\epsfxsize=3.5in
\centerline{\epsffile{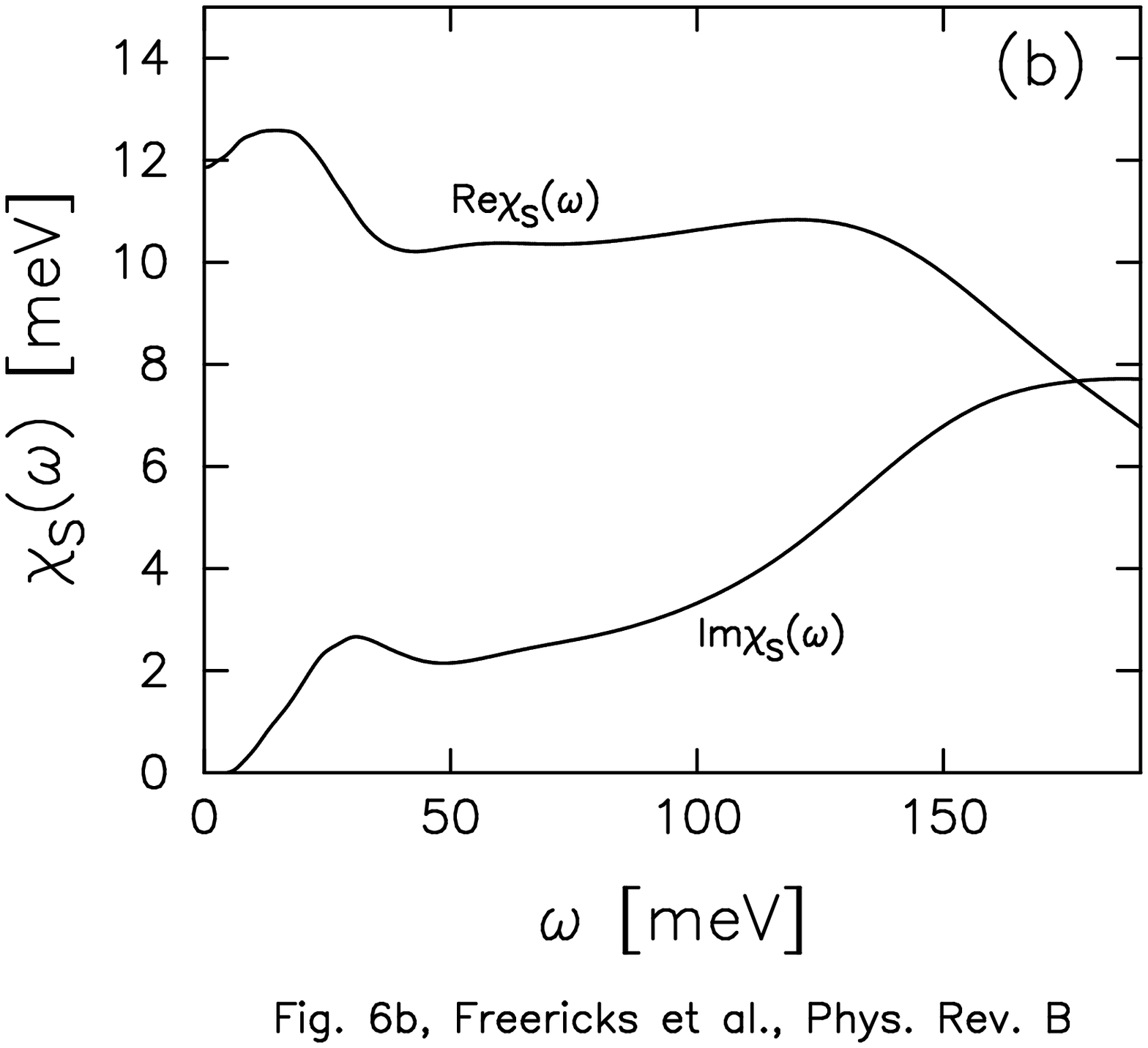}}
\caption{
Self-energy function $\chi(\omega)$ plotted in the (a) normal and (b)
superconducting states.  
These functions are nearly identical, except for small differences at
the lowest energies.
}
\label{fig: chi}
\end{figure}

In Fig.~\ref{fig: chi}, we plot the real and imaginary
parts of $\chi$ for (a) the normal state and (b) the superconducting state.
This function vanishes for the case of a constant density of states.
The chemical potentials are $\mu_N=15.70$~meV and $\mu_S=15.97$~meV
in the normal and superconducting states.  The value of the real part of
$\chi$ is of this order of magnitude.  The 
normal and superconducting self-energies differ only at the lowest energies.

\begin{figure}[htbf]
\epsfxsize=3.5in
\centerline{\epsffile{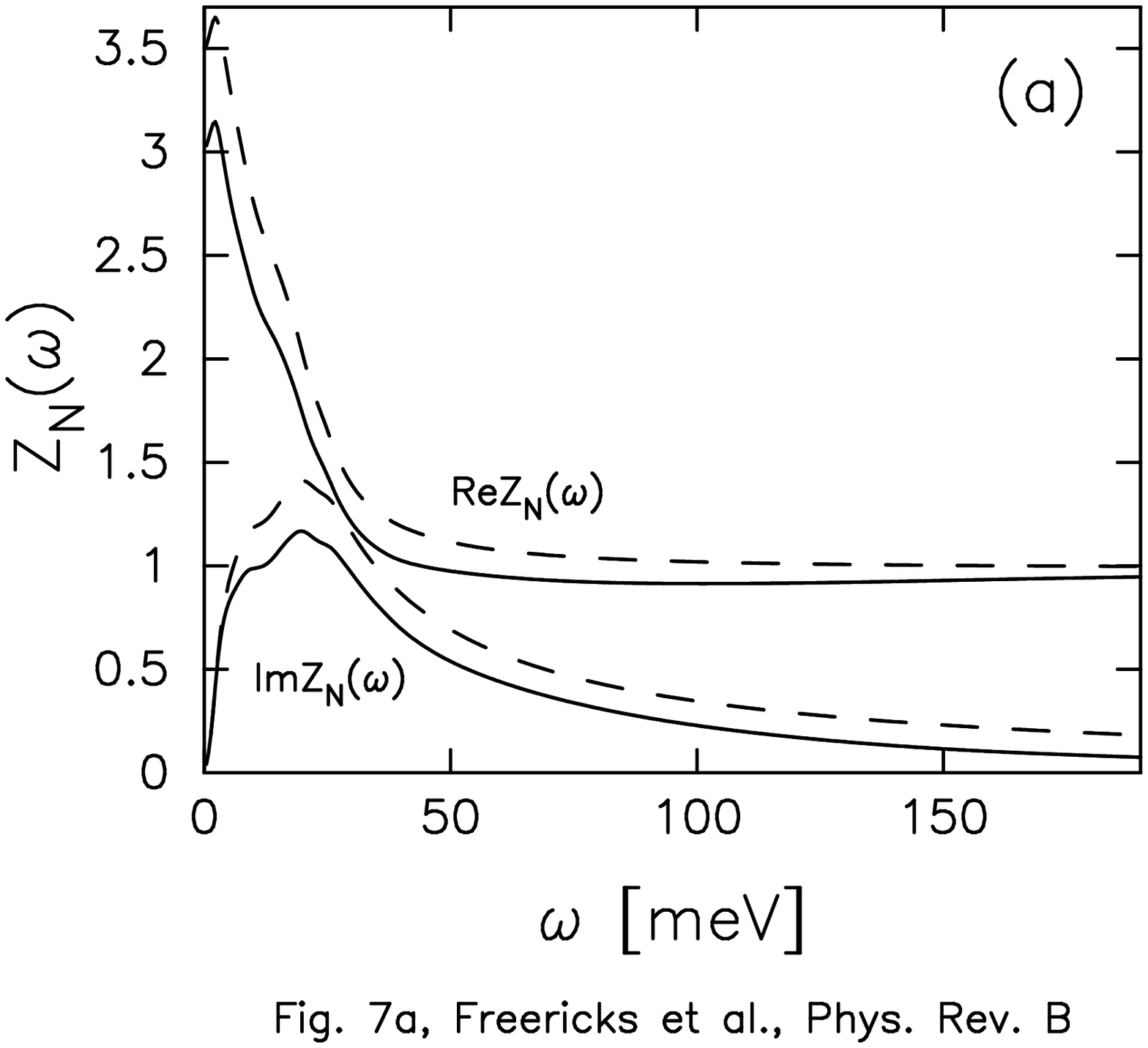}}
\epsfxsize=3.5in
\centerline{\epsffile{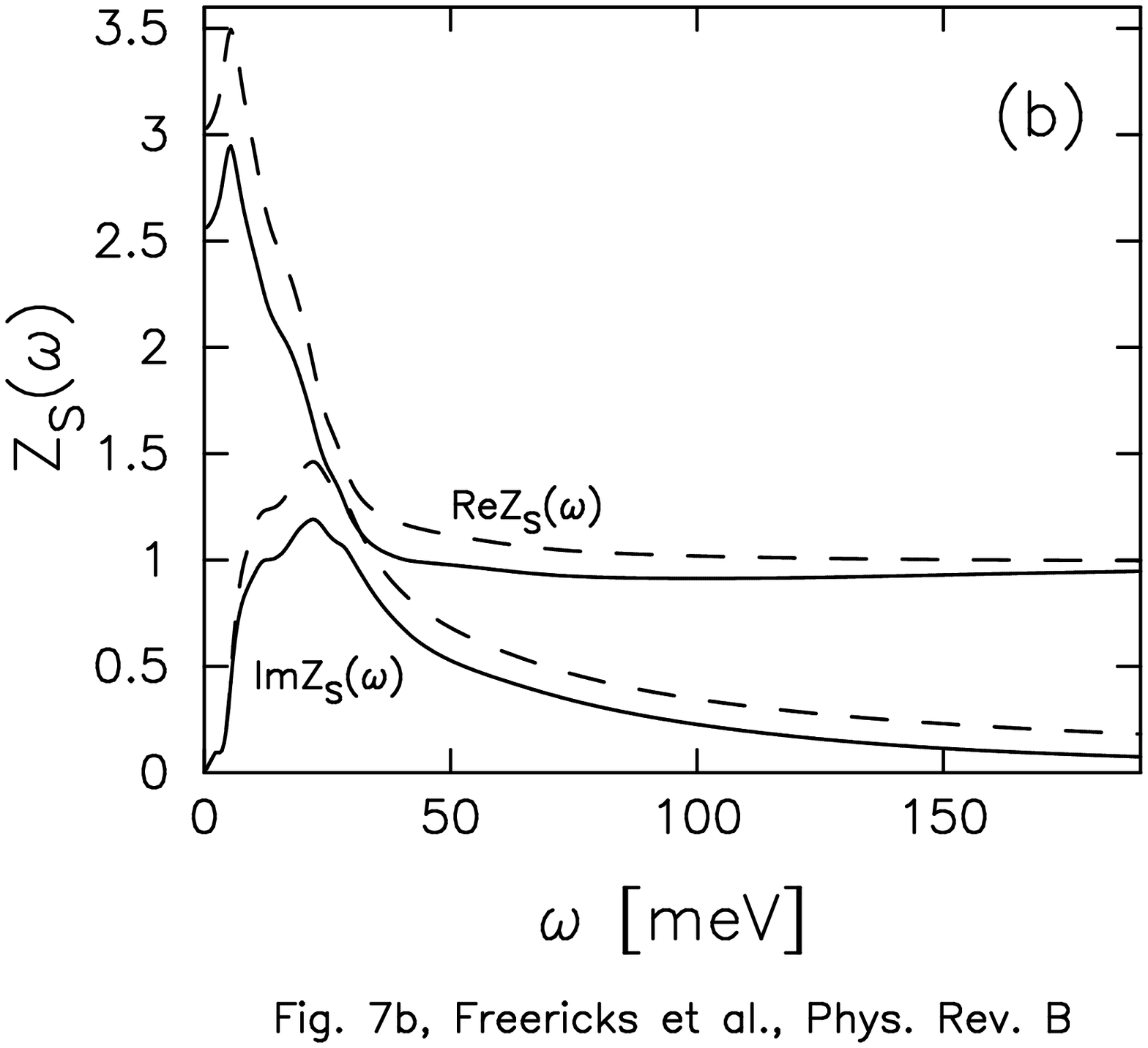}}
\caption{
The renormalization function for (a) the normal state and (b)
the superconducting state.  The solid curves are for a nonconstant
density of states and the dashed curves are for a constant density of
states.  Note that ${\rm Re}Z$ can dip below 1 for the nonconstant density
of states case.
}
\label{fig: z}
\end{figure}

The renormalization function $Z(\omega)$ is plotted in 
Fig.~\ref{fig: z}, for (a) the normal phase and (b) the superconducting
phase.  The solid lines are for the nonconstant density of states case and the
dashed lines are for the constant density of states case. 
Note how in the nonconstant
density of states case, the effective strength of the electron phonon 
coupling, measured by $Z_N(0)-1$ is closer to 2.1 than the value of
$\lambda$, which is 2.7.  It is the former value that is the true measure
of the electron-phonon coupling strength with a nonconstant density
of states.  Note that the main differences between the nonconstant
and constant density of states calculations is that the overall scale
is larger for the latter.  These functions vary from the normal to the 
superconducting state only at low energies as expected.  Note further, that
the real part of $Z$ can dip below 1 for the nonconstant density of states
case (it never does for a constant density of states).

\begin{figure}[htbf]
\epsfxsize=3.5in
\centerline{\epsffile{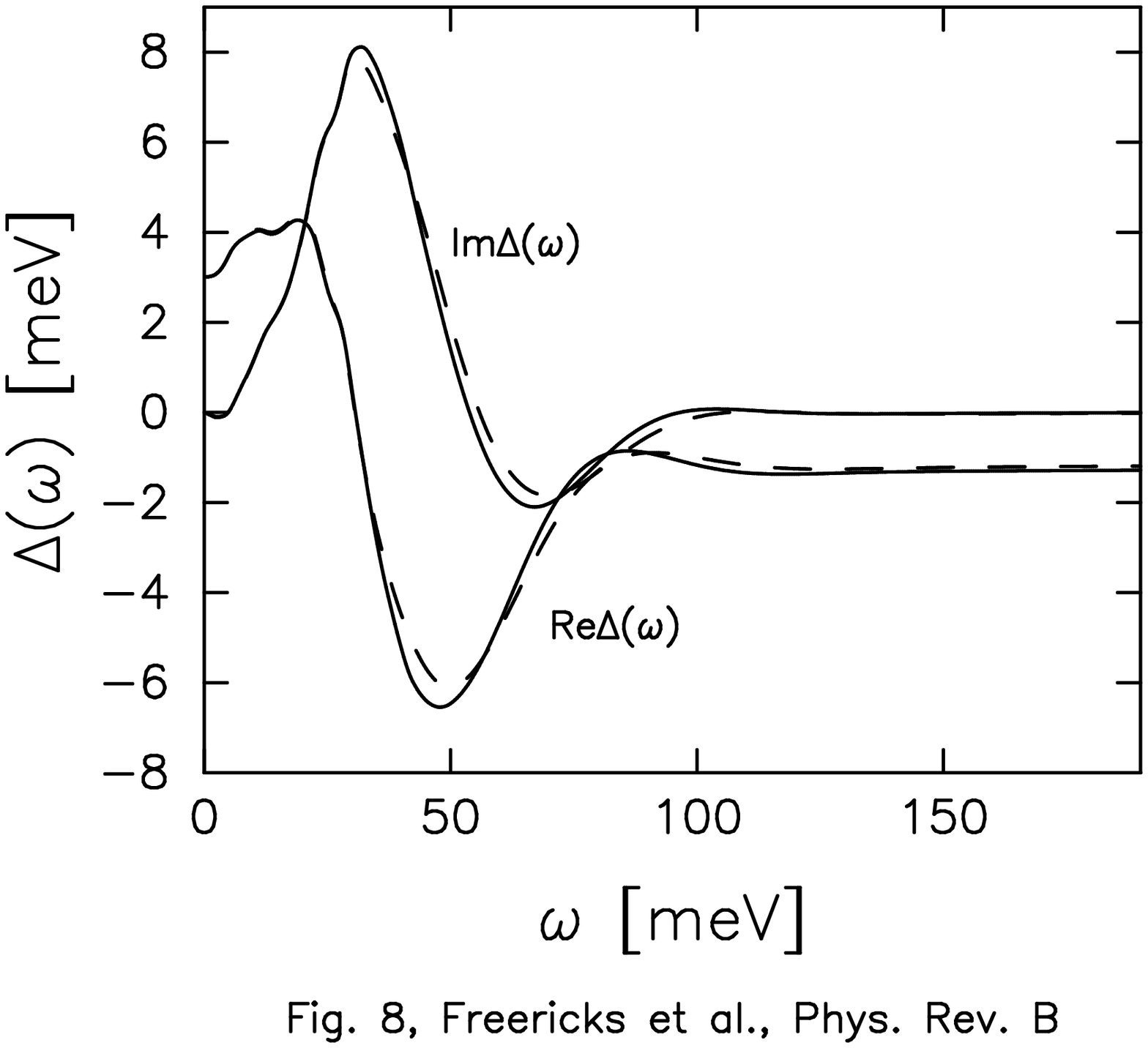}}
\caption{
Superconducting gap function for (solid) nonconstant density of states
and (dashed) constant density of states.  The two curves are
nearly identical at low and high energies, but differ from about 30~meV
to 130~meV.  The sharp overswing in the experimental RDOS data is 
better reproduced by the 
more rapid formation of the high-energy peak (dip) in the real 
(imaginary) part of the nonconstant-density-of-states gap function.
}
\label{fig: delta}
\end{figure}

Finally, the superconducting gap function $\Delta(\omega)$ is shown 
in Fig.~\ref{fig: delta}.  The results for the nonconstant density of states
(solid line) and the constant density of states (dashed line) are
nearly identical at both low and high energies.  In the range from 30~meV to 
130~meV, the curves deviate from each other: in the nonconstant
density of states case, the peak (real part) and dip (imaginary part)
form faster than in the constant density of states case. This is what 
produces the sharper overswing in the RDOS for the
nonconstant density of states calculation. 

\section{Conclusions}

We have performed a modified McMillan-Rowell tunneling inversion including
the effects of the nonconstant electronic density of states near the
Fermi level 
for Nb$_3$Sn.  The Eliashberg function that 
is found by this inversion has a larger value of $\lambda$ than generally
accepted for this material, but the effective value of $\lambda$ derived
from the renormalization function is more reasonable.  Our extracted 
$\alpha^2F(\Omega)$ also has a number of expected features including large
peaks at about 6 and 8~meV representing the soft phonon modes, and a sharply
peaked $\alpha^2(\Omega)$ below 10~meV which agrees with the
theoretical predictions.  Even taking
into account the energy dependence of the density of states, 
we are unable to produce excellent fits of the tunneling
data, though we can better reproduce the overswing observed at high
frequencies. 
We believe the problem is partially numerical, as the tunneling inversion  
tries to 
force sharp spikes into $\alpha^2F$, but the fit is still too
poor at the lowest energies  and in the overswing region
to say that properly including the
energy dependence of the density of states is enough to completely understand 
the tunneling data. It may be that the tunneling is modified by anisotropic
effects, by a 
narrow proximity-coupled layer or by other effects such as anharmonicity. 

\acknowledgments
We would like to acknowledge stimulating discussions with M. Beasley, J. 
Carbotte, 
B. Klein, B. Mitrovi\'c, E. Nicol, and D. Rudman. We also want to thank
D. Rudman for sharing his data with us.
This work was supported by the National Science Foundation
under grant DMR-9973225.


\begin{table}
\caption{Calculated properties of the Eliashberg function extracted
from the tunneling inversion. }
\begin{tabular}{lccccccc}
\tableline
inversion & $\lambda$ & $\mu^*$ & $\omega_{\rm ln}$ & $A$ & $T_c$ &
error & error\\
DOS & & & [meV] & [meV] & [K] & max. & r.m.s.\\
\tableline
\noalign{\vskip0.03in}
nonconstant & 2.738 & 0.286 & 7.072 & 13.652 & 19 & 0.033 & 0.0048 \\
constant &  2.501 & 0.210 & 6.415 & 11.130 & 23 & 0.034 & 0.0055 \\
\tableline
\end{tabular}
\label{table: a2f}
\end{table}

\end{document}